\documentstyle[12pt,axodraw]{article}
\title{Anomalous $W$ boson production at HERA}

\author{
         M.N.Dubinin \\
       {\small \it Institute of Nuclear Physics, Moscow State University} \\
       {\small \it  119899 Moscow, Russia} \\
         H.S.Song \\
       {\small \it Center for Theoretical Physics,} 
       {\small \it Seoul National University} \\
       {\small \it Seoul, 151-742, Korea} } 
\date{}

\begin{document}
\maketitle

\begin{abstract}
We present the results of complete tree level calculation for $W$ boson
production processes $e^- p \rightarrow e^- \mu^+ \nu_{\mu} X$ and
$e^- p \rightarrow e^- \mu^- \bar \nu_{\mu} X$ introducing anomalous 
$WW\gamma$ and $WWZ$ couplings. Detailed results for the distributions
of final state particles are obtained. In the region of small momentum
transfer we calculate the contribution of hadronlike photon component
in the structure function approach. 
\end{abstract}

\section{Introduction}

In recent years the charged and neutral current sectors of the Standard
Model have been tested with excellent precision in the experiments at
LEP and SLC. However the gauge boson sector still remains practically
untouched by direct measurements of high accuracy. Deviations of three
and four gauge boson couplings from the Standard Model values would be
an obvious signal of some new physics.

At present time the best limits on anomalous three vector boson
couplings are given by CDF and D0 data (Fermilab Tevatron, $\sqrt{s}=$
1.8 TeV) \cite{Tevatron}. From the measurement of $W \gamma$ and $WW$
production these collaborations set the limits of order 1 on the
deviations of ($k$, $\lambda$) couplings (see section 2.1) from the
Standard Model values. Significant improvement of these limits (one
order of magnitude) will be achieved by the detection of $WW$ production
at LEP2 \cite{LEP2}.

In this paper we consider the possibilities of HERA $ep$ collider
(30 GeV electrons on 820 GeV protons, $\sqrt{s}=$ 314 GeV)
for the measurement of the vector boson anomalous couplings. 
The main difference between our study and the previous investigations
is the exact calculation of tree level amplitude for complete set of diagrams
with the four particle final state, including nonstandard $WW\gamma$ and
$WWZ$ vertices.
At present time the luminosity of HERA (several $pb^{-1}$/year) is too small
to produce sufficient number of $W$ bosons. However, after the luminosity
upgrade to 100 - 200 $pb^{-1}$ the detection of anomalous signal or
setting new limits on the anomalous couplings becomes realistic.

\section{The reactions $e^- p \rightarrow
                  e^- \mu^+ \nu_{\mu} X$, $e^- \mu^- \bar \nu_{\mu} X$}

It is known from the previous study (in particular we would like to
distinguish the paper \cite{BVZ}) that in the Standard Model $W$-bosons
are produced in $ep$ scattering mainly in the channels $ep \rightarrow
eWX$. The contribution of the channel with neutrino in the final state
$ep \rightarrow \nu_e W^-X$ is 20 times smaller. The following decay of
$W$ boson to muon and muonic neutrino produces the
four-fermion state  $e^- \mu \nu_{\mu} q$, and the corresponding
event signature is muon(antimuon) with missing transverse
momentum. The signal of $W$ boson production in leptonic channels can
be observed easier than in hadronic $W$ decay channels, where
large QCD background processes must be carefully separated from the
signal.

Ten Feynman diagrams for the reaction $e^- q_1 \rightarrow e^- \mu^+
\nu_{\mu} q_2$ are shown in Fig.1. All diagrams have intermediate $W$ and
the properties of final state are defined by $W$ interaction dynamics.
In this sense all diagrams are '$W^+$ producing' and there are no irreducible
background graphs that could be neglected in order to simplify the procedure.
If we replace diagram 5 by similar one where $W^-$ boson is radiated from
the initial electron and change $\mu^+ \nu_{\mu}$ to $\mu^- \bar \nu_{\mu}$,
$q_{1,2} \rightarrow \bar q_{1,2}$, 
we obtain a set of ten diagrams for the process $e^- p \rightarrow 
e^- \mu^- \bar \nu_{\mu} X$ which is '$W^-$ producing'.  
Diagrams 4 and 10 in both cases contain $WW\gamma$ and $WWZ$ vertices 
and in the following we shall use for them nonstandard gauge invariant 
structure. 

If we separate subsets of diagrams from the complete tree level set in
Fig.1 and then separate Feynman subgraphs from these subsets, we 
obtain some approximations that were used in the previous calculations of
$W$ production processes. The simplest approximation is given by diagrams
2,3,4 where $t$-channel photon and s-channel $W$ are taken on-shell
($\gamma^* q_1 \rightarrow W^* q_2$, $2 \rightarrow 2$ subprocess
approximation). If we integrate then with equivalent photon structure 
function for incoming $\gamma^*$ and consider the decay of on-shell $W$
to $\mu \nu_{\mu}$, rather satisfactory estimate of total cross section 
can be obtained. The calculation in $\gamma^* q_1 \rightarrow W^* q_2$
subprocess approximation with anomalous vector boson couplings can be found
in \cite{KLS}. Weak correction to this result is given by 
diagrams 8,9,10
(containing subprocess $Z^* q_1 \rightarrow W^+ q_2$). In the case if the
accuracy of
equivalent photon approximation is not sufficient, at the next step one 
could consider $e^- q_1 \rightarrow e^- W^+ q_2$, i.e. $2 \rightarrow 3$ 
process 
approximation with on-shell vector boson. Complete tree level 
calculation for the process $e^- q_1 \rightarrow e^- W^+ q_2$
by means of helicity amplitude method was performed in \cite{BZ}.
If we take the 
amplitude with
$W$ off-shell decaying to fermionic pair (for instance, $e^- q_1 \rightarrow 
e^- \mu^+ \nu_{\mu} q_2$ in the $2 \rightarrow 4$ process approximation), 
the subset of diagrams (2,3,4,5,8,9,10) becomes gauge noninvariant and in
order to restore the gauge invariance it is necessary to add ladder
diagrams 1,6,7. In the papers \cite{BVZ,other} complete tree
level calculation was performed for the case of
Standard Model. We generalize these calculations in the case of 
complete tree level
$2 \rightarrow 4$ muonic channels with
anomalous C and P conserving three vector boson couplings, and
compare some of our results with \cite{noyes} where similar analysis
was done by means of EPVEC generator \cite{BVZ}.  

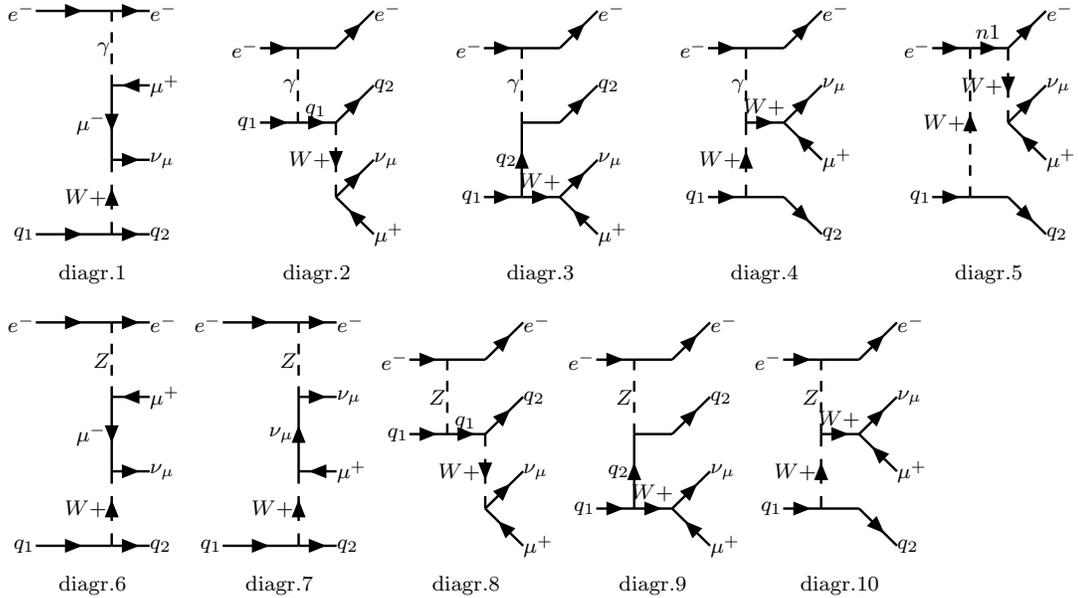
\begin{figure}[h]
{\def\chepscale{1.3} 
\unitlength=\chepscale pt
\SetWidth{0.7}      
\SetScale{\chepscale}
\scriptsize    
\begin{picture}(50,90)(0,0)
\Text(8.2,77.8)[r]{$e^-$}
\ArrowLine(8.5,77.8)(30.5,77.8) 
\Text(41.8,77.8)[l]{$e^-$}
\ArrowLine(30.5,77.8)(41.5,77.8) 
\Text(30.1,67.0)[r]{$\gamma$}
\DashLine(30.5,77.8)(30.5,56.2){3.0} 
\Text(41.8,56.2)[l]{$\mu^+$}
\ArrowLine(41.5,56.2)(30.5,56.2) 
\Text(29.1,45.4)[r]{$\mu^-$}
\ArrowLine(30.5,56.2)(30.5,34.6) 
\Text(41.8,34.6)[l]{$\nu_{\mu}$}
\ArrowLine(30.5,34.6)(41.5,34.6) 
\Text(29.1,23.8)[r]{$W+$}
\DashArrowLine(30.5,13.0)(30.5,34.6){3.0} 
\Text(8.2,13.0)[r]{$q_1$}
\ArrowLine(8.5,13.0)(30.5,13.0) 
\Text(41.8,13.0)[l]{$q_2$}
\ArrowLine(30.5,13.0)(41.5,13.0) 
\Text(25,0)[b] {diagr.1}
\end{picture} \ 
\begin{picture}(50,90)(0,0)
\Text(8.2,67.0)[r]{$e^-$}
\ArrowLine(8.5,67.0)(19.5,67.0) 
\Line(19.5,67.0)(30.5,67.0) 
\Text(41.8,77.8)[l]{$e^-$}
\ArrowLine(30.5,67.0)(41.5,77.8) 
\Text(19.1,56.2)[r]{$\gamma$}
\DashLine(19.5,67.0)(19.5,45.4){3.0} 
\Text(8.2,45.4)[r]{$q_1$}
\ArrowLine(8.5,45.4)(19.5,45.4) 
\Text(24.8,48.2)[b]{$q_1$}
\ArrowLine(19.5,45.4)(30.5,45.4) 
\Text(41.8,56.2)[l]{$q_2$}
\ArrowLine(30.5,45.4)(41.5,56.2) 
\Text(29.1,34.6)[r]{$W+$}
\DashArrowLine(30.5,45.4)(30.5,23.8){3.0} 
\Text(41.8,34.6)[l]{$\nu_{\mu}$}
\ArrowLine(30.5,23.8)(41.5,34.6) 
\Text(41.8,13.0)[l]{$\mu^+$}
\ArrowLine(41.5,13.0)(30.5,23.8) 
\Text(25,0)[b] {diagr.2}
\end{picture} \ 
\begin{picture}(50,90)(0,0)
\Text(8.2,67.0)[r]{$e^-$}
\ArrowLine(8.5,67.0)(19.5,67.0) 
\Line(19.5,67.0)(30.5,67.0) 
\Text(41.8,77.8)[l]{$e^-$}
\ArrowLine(30.5,67.0)(41.5,77.8) 
\Text(19.1,56.2)[r]{$\gamma$}
\DashLine(19.5,67.0)(19.5,45.4){3.0} 
\Line(19.5,45.4)(30.5,45.4) 
\Text(41.8,56.2)[l]{$q_2$}
\ArrowLine(30.5,45.4)(41.5,56.2) 
\Text(18.1,34.6)[r]{$q_2$}
\ArrowLine(19.5,23.8)(19.5,45.4) 
\Text(8.2,23.8)[r]{$q_1$}
\ArrowLine(8.5,23.8)(19.5,23.8) 
\Text(24.8,26.6)[b]{$W+$}
\DashArrowLine(19.5,23.8)(30.5,23.8){3.0} 
\Text(41.8,34.6)[l]{$\nu_{\mu}$}
\ArrowLine(30.5,23.8)(41.5,34.6) 
\Text(41.8,13.0)[l]{$\mu^+$}
\ArrowLine(41.5,13.0)(30.5,23.8) 
\Text(25,0)[b] {diagr.3}
\end{picture} \ 
\begin{picture}(50,90)(0,0)
\Text(8.2,67.0)[r]{$e^-$}
\ArrowLine(8.5,67.0)(19.5,67.0) 
\Line(19.5,67.0)(30.5,67.0) 
\Text(41.8,77.8)[l]{$e^-$}
\ArrowLine(30.5,67.0)(41.5,77.8) 
\Text(19.1,56.2)[r]{$\gamma$}
\DashLine(19.5,67.0)(19.5,45.4){3.0} 
\Text(24.8,48.2)[b]{$W+$}
\DashArrowLine(19.5,45.4)(30.5,45.4){3.0} 
\Text(41.8,56.2)[l]{$\nu_{\mu}$}
\ArrowLine(30.5,45.4)(41.5,56.2) 
\Text(41.8,34.6)[l]{$\mu^+$}
\ArrowLine(41.5,34.6)(30.5,45.4) 
\Text(18.1,34.6)[r]{$W+$}
\DashArrowLine(19.5,23.8)(19.5,45.4){3.0} 
\Text(8.2,23.8)[r]{$q_1$}
\ArrowLine(8.5,23.8)(19.5,23.8) 
\Line(19.5,23.8)(30.5,23.8) 
\Text(41.8,13.0)[l]{$q_2$}
\ArrowLine(30.5,23.8)(41.5,13.0) 
\Text(25,0)[b] {diagr.4}
\end{picture} \ 
\begin{picture}(50,90)(0,0)
\Text(8.2,67.0)[r]{$e^-$}
\ArrowLine(8.5,67.0)(19.5,67.0) 
\Text(24.8,69.8)[b]{$n1$}
\ArrowLine(19.5,67.0)(30.5,67.0) 
\Text(41.8,77.8)[l]{$e^-$}
\ArrowLine(30.5,67.0)(41.5,77.8) 
\Text(29.1,56.2)[r]{$W+$}
\DashArrowLine(30.5,67.0)(30.5,45.4){3.0} 
\Text(41.8,56.2)[l]{$\nu_{\mu}$}
\ArrowLine(30.5,45.4)(41.5,56.2) 
\Text(41.8,34.6)[l]{$\mu^+$}
\ArrowLine(41.5,34.6)(30.5,45.4) 
\Text(18.1,45.4)[r]{$W+$}
\DashArrowLine(19.5,23.8)(19.5,67.0){3.0} 
\Text(8.2,23.8)[r]{$q_1$}
\ArrowLine(8.5,23.8)(19.5,23.8) 
\Line(19.5,23.8)(30.5,23.8) 
\Text(41.8,13.0)[l]{$q_2$}
\ArrowLine(30.5,23.8)(41.5,13.0) 
\Text(25,0)[b] {diagr.5}
\end{picture} \ 
\begin{picture}(50,90)(0,0)
\Text(8.2,77.8)[r]{$e^-$}
\ArrowLine(8.5,77.8)(30.5,77.8) 
\Text(41.8,77.8)[l]{$e^-$}
\ArrowLine(30.5,77.8)(41.5,77.8) 
\Text(30.1,67.0)[r]{$Z$}
\DashLine(30.5,77.8)(30.5,56.2){3.0} 
\Text(41.8,56.2)[l]{$\mu^+$}
\ArrowLine(41.5,56.2)(30.5,56.2) 
\Text(29.1,45.4)[r]{$\mu^-$}
\ArrowLine(30.5,56.2)(30.5,34.6) 
\Text(41.8,34.6)[l]{$\nu_{\mu}$}
\ArrowLine(30.5,34.6)(41.5,34.6) 
\Text(29.1,23.8)[r]{$W+$}
\DashArrowLine(30.5,13.0)(30.5,34.6){3.0} 
\Text(8.2,13.0)[r]{$q_1$}
\ArrowLine(8.5,13.0)(30.5,13.0) 
\Text(41.8,13.0)[l]{$q_2$}
\ArrowLine(30.5,13.0)(41.5,13.0) 
\Text(25,0)[b] {diagr.6}
\end{picture} \ 
\begin{picture}(50,90)(0,0)
\Text(8.2,77.8)[r]{$e^-$}
\ArrowLine(8.5,77.8)(30.5,77.8) 
\Text(41.8,77.8)[l]{$e^-$}
\ArrowLine(30.5,77.8)(41.5,77.8) 
\Text(30.1,67.0)[r]{$Z$}
\DashLine(30.5,77.8)(30.5,56.2){3.0} 
\Text(41.8,56.2)[l]{$\nu_{\mu}$}
\ArrowLine(30.5,56.2)(41.5,56.2) 
\Text(29.1,45.4)[r]{$\nu_{\mu}$}
\ArrowLine(30.5,34.6)(30.5,56.2) 
\Text(41.8,34.6)[l]{$\mu^+$}
\ArrowLine(41.5,34.6)(30.5,34.6) 
\Text(29.1,23.8)[r]{$W+$}
\DashArrowLine(30.5,13.0)(30.5,34.6){3.0} 
\Text(8.2,13.0)[r]{$q_1$}
\ArrowLine(8.5,13.0)(30.5,13.0) 
\Text(41.8,13.0)[l]{$q_2$}
\ArrowLine(30.5,13.0)(41.5,13.0) 
\Text(25,0)[b] {diagr.7}
\end{picture} \ 
\begin{picture}(50,90)(0,0)
\Text(8.2,67.0)[r]{$e^-$}
\ArrowLine(8.5,67.0)(19.5,67.0) 
\Line(19.5,67.0)(30.5,67.0) 
\Text(41.8,77.8)[l]{$e^-$}
\ArrowLine(30.5,67.0)(41.5,77.8) 
\Text(19.1,56.2)[r]{$Z$}
\DashLine(19.5,67.0)(19.5,45.4){3.0} 
\Text(8.2,45.4)[r]{$q_1$}
\ArrowLine(8.5,45.4)(19.5,45.4) 
\Text(24.8,48.2)[b]{$q_1$}
\ArrowLine(19.5,45.4)(30.5,45.4) 
\Text(41.8,56.2)[l]{$q_2$}
\ArrowLine(30.5,45.4)(41.5,56.2) 
\Text(29.1,34.6)[r]{$W+$}
\DashArrowLine(30.5,45.4)(30.5,23.8){3.0} 
\Text(41.8,34.6)[l]{$\nu_{\mu}$}
\ArrowLine(30.5,23.8)(41.5,34.6) 
\Text(41.8,13.0)[l]{$\mu^+$}
\ArrowLine(41.5,13.0)(30.5,23.8) 
\Text(25,0)[b] {diagr.8}
\end{picture} \ 
\begin{picture}(50,90)(0,0)
\Text(8.2,67.0)[r]{$e^-$}
\ArrowLine(8.5,67.0)(19.5,67.0) 
\Line(19.5,67.0)(30.5,67.0) 
\Text(41.8,77.8)[l]{$e^-$}
\ArrowLine(30.5,67.0)(41.5,77.8) 
\Text(19.1,56.2)[r]{$Z$}
\DashLine(19.5,67.0)(19.5,45.4){3.0} 
\Line(19.5,45.4)(30.5,45.4) 
\Text(41.8,56.2)[l]{$q_2$}
\ArrowLine(30.5,45.4)(41.5,56.2) 
\Text(18.1,34.6)[r]{$q_2$}
\ArrowLine(19.5,23.8)(19.5,45.4) 
\Text(8.2,23.8)[r]{$q_1$}
\ArrowLine(8.5,23.8)(19.5,23.8) 
\Text(24.8,26.6)[b]{$W+$}
\DashArrowLine(19.5,23.8)(30.5,23.8){3.0} 
\Text(41.8,34.6)[l]{$\nu_{\mu}$}
\ArrowLine(30.5,23.8)(41.5,34.6) 
\Text(41.8,13.0)[l]{$\mu^+$}
\ArrowLine(41.5,13.0)(30.5,23.8) 
\Text(25,0)[b] {diagr.9}
\end{picture} \ 
\begin{picture}(50,90)(0,0)
\Text(8.2,67.0)[r]{$e^-$}
\ArrowLine(8.5,67.0)(19.5,67.0) 
\Line(19.5,67.0)(30.5,67.0) 
\Text(41.8,77.8)[l]{$e^-$}
\ArrowLine(30.5,67.0)(41.5,77.8) 
\Text(19.1,56.2)[r]{$Z$}
\DashLine(19.5,67.0)(19.5,45.4){3.0} 
\Text(24.8,48.2)[b]{$W+$}
\DashArrowLine(19.5,45.4)(30.5,45.4){3.0} 
\Text(41.8,56.2)[l]{$\nu_{\mu}$}
\ArrowLine(30.5,45.4)(41.5,56.2) 
\Text(41.8,34.6)[l]{$\mu^+$}
\ArrowLine(41.5,34.6)(30.5,45.4) 
\Text(18.1,34.6)[r]{$W+$}
\DashArrowLine(19.5,23.8)(19.5,45.4){3.0} 
\Text(8.2,23.8)[r]{$q_1$}
\ArrowLine(8.5,23.8)(19.5,23.8) 
\Line(19.5,23.8)(30.5,23.8) 
\Text(41.8,13.0)[l]{$q_2$}
\ArrowLine(30.5,23.8)(41.5,13.0) 
\Text(25,0)[b] {diagr.10}
\end{picture} \ 
}
\caption{Feynman diagrams for the process $e^- q_1 \rightarrow
         e^- \mu^+ \nu_{\mu} q_2$}
\end{figure}

While the results for total rate provided by $2 \rightarrow 2$ 
and $2 \rightarrow 3$ approximations can be quite satisfactory, it is
not possible to calculate precisely most of the distributions of experimental
interest. The accuracy of equivalent photon approximation becomes rather
poor (see section 4) especially at large transverse momenta and the 
narrow-width approximation for
the $W$ is usually not good near the $W$ production threshold. At the same time
it is obviously difficult to make any conclusions about the origin of new 
phenomena observing only the deviation of event counting rate from the Standard
Model value. It is important to know what regions of phase space are 
affected by new interaction dynamics and what is the ratio new signal/
background, i.e. to calculate precisely the distributions of particles in the 
final state. 

Largest contribution to the cross section of $e^- q_1 \rightarrow
e^- \mu^+ \nu_{\mu} q_2$ process is given by diagram 3 when photon
and t-channel quark are close to mass shell \cite{resolved}. In this 
configuration QCD corrections become large and $t$-channel intermediate
quark can appear nonperturbatively as a constituent of photon ('resolved
photon' contribution) when it is usually described by experimentally measured 
gamma structure function. The process of $W$ production in this picture is
quark-antiquark $q_{\gamma} q_p$ fusion to $W$ when photon fragments 
into quark constituents $q_{\gamma}$ before interacting with the 
proton constituent quark $q_p$. It was shown in \cite{BVZ},\cite{resolved} 
that resolved
photon mechanism is not dominant, but nevertheless the region of small $t$ 
requires special consideration and careful separation of 'resolved' and 
ordinary contributions is needed. To be sure that the numbers are not
changed significantly by the new parametrizations of $\gamma$ and $p$
structure functions,
we repeat the cross section calculation of the resolved part using the 
scheme similar to one proposed in \cite{KLS}.
    
\subsection{Anomalous three vector boson couplings}

General effective lagrangian of two charged and one
neutral gauge boson interaction was proposed in \cite{GG}. The
restrictions on the lagrangian imposed by invariance under
discrete symmetries and gauge invariance were considered in \cite{HPZH}. 
U(1) gauge invariant, C and P parity conserving effective lagrangian has 
the form   
      
\begin{equation}
L_{eff}=g_V (W^+_{\mu \nu} W^{\mu} V^{\nu}
            -W^{\mu \nu} W^+_{\mu} V_{\nu}
            +k \; W^+_{\mu} W_{\nu} V^{\mu \nu}
  +\frac{\lambda}{m^2_W} W^+_{\rho \mu} W^{\mu}_{\; \; \nu} V^{\nu \rho}) 
\end{equation}
where $g_{\gamma}=e$ and $g_Z=e cos\vartheta_W/sin\vartheta_W$,
$W_{\mu \nu}=\partial_{\mu} W_{\nu}- \partial_{\nu} W_{\mu}$, 
$V_{\mu \nu}=\partial_{\mu} V_{\nu}- \partial_{\nu} V_{\mu}$. 
Spatial structure of the fourth term in the 
lagrangian of dimension six is multiplied by $m_W^{-2}$ factor, so 
parameters $\lambda$ and $k$ are dimensionless. In the momentum space 
if all momenta are incoming ($p_1+p_2+p_3=0$), we have the following
expression for $W^+(p_1)W^-(p_2)V(p_3)$ vertex

\begin{eqnarray}
\Gamma_{\mu \nu \rho} (p_1,p_2,p_3) 
                          = g_V [g_{\mu \nu} (p_1-p_2-\frac{\lambda}{m^2_W}
                               ((p_2p_3)p_1-(p_1p_3)p_2)_{\rho} \\ \nonumber 
                            + g_{\mu \rho} (kp_3-p_1+\frac{\lambda}{m^2_W}
                               ((p_2p_3)p_1-(p_1p_2)p_3)_{\nu} \\ \nonumber
                            + g_{\nu \rho} (p_2-kp_3-\frac{\lambda}{m^2_W}
                               ((p_1p_3)p_2-(p_1p_2)p_3)_{\mu} \\ \nonumber
 +\frac{\lambda}{m^2_W}({p_2}_{\mu} {p_3}_{\nu} {p_1}_{\rho}
                             - {p_3}_{\mu} {p_1}_{\nu} {p_2}_{\rho})]
\end{eqnarray}

In the special case $\lambda=0$, $k=1$ this vertex reduces to Standard
Model one.

\section{Resolved photon contribution}

In this section we follow the scheme proposed in \cite{KLS} for
the calculation of resolved photon contribution (diagram 3, Fig.1),
but in our case $W$ boson is off-shell and consequently the scale
of the equivalent photon approximation is different.
Similar procedure (including the corrections from 'finite terms') 
was considered in \cite{BVZ}. We separate the
'resolved' and 'direct' production mechanisms at the scale $\Lambda^2=
(q_1-q_W)^2$ ($q_1$ and $q_W$ are the momenta of initial quark and
intermediate $W$, correspondingly):
\begin{equation}
\sigma=\sigma_{resolved}+\int^{-\Lambda^2}\frac{d\sigma_{dir}}{dt}dt
\end{equation}
Second term in this formula ('direct') is calculated numerically using
the exact matrix element for 10 diagrams (see Fig.1) with the transferred
momentum cutoff at $t= \; \Lambda^2$.
Resolved photon cross section in the case of monoenergetic initial gamma
on shell is given by the convolution of photon
structure function (measured in $\gamma \gamma$ collisions) and 
$q_{\gamma} q_p \rightarrow W$ fusion cross section
\begin{equation}
\sigma_{resolved}=\int dx_1 dx_2
                     f_{q_1 / {\gamma}}(x_1,Q^2_{\gamma}) \; 
                       \sigma(q_1 \bar q_2 \rightarrow W) \;
                    f_{q_2/p} (x_2, Q^2_p)
\end{equation}
$Q^2_{\gamma}$ and $Q^2_p$ are the scales for photon and proton structure 
functions, correspondingly.
Using the Breit-Wigner formula in the approximation of infinitely
small $W$ width we get
\begin{equation}
\sigma(q_1 \bar q_2 \rightarrow W)=\frac{\pi \sqrt{2}}{3} G_F m^2_W
                                    |V_{12}|^2 \delta(x_1 x_2 s-m^2_W)
\end{equation}
where $V_{12}$ is the CKM matrix element for charged current
$\psi_{q_1}\psi_{q_2}$; here and in the following we are not indicating
the sum over possible quark spieces.
The experimentally measured photon structure function $f_{q_1 / {\gamma}}$
includes point-like
as well as hadron-like parts. In the leading logarithmic approximation
the perturbative (point-like) part of photon structure function can be
expressed as
\begin{equation}
                  f^{LO}_{q_1 / {\gamma}}(x_1,Q^2_{\gamma})=
 \frac{3 \alpha e^2_q}{2 \pi} [x^2+(1-x)^2]log\frac{Q^2_{\gamma}}{\Lambda^2}
\end{equation}
where $e_q$ is the quark charge.
This part was already taken into account by our calculation for direct
contribution in (3). In order to avoid double counting in the contributions
from gamma structure function and from direct process it is necessary to 
subtract the point-like term (6) from $f_{q_1 / {\gamma}}(x_1,Q^2_{\gamma})$. 
This procedure was illustrated explicitly for the more simple example of
the reaction $\gamma q_1 \rightarrow V q_2$ ($V=Z,W$) in \cite{KLS} 
and it was shown that in the case of $Z$ production indeed the LO counterterm 
(6) rescales the leading logarithmic structure in the direct part from
$\Lambda^2$ to $Q^2_{\gamma}$ when $f_{q_1 / {\gamma}}(x_1,Q^2_{\gamma})$ 
is taken at momentum transfer $Q^2_{\gamma}$.
Finally the dependence from $\Lambda^2$ is absent in the sum of resolved 
and direct contributions. Our case is certainly more complicated but
double counting can be avoided at least on the leading logarithmic level.

Introducing the usual equivalent photon approximation \cite{WWA} 
for gamma in the initial state
\begin{equation}
f_{q/e}(x,Q^2_{\gamma},Q^2_{WW})=\int^1_x \; \frac{dy}{y} 
                             f_{q_1/{\gamma}} (\frac{x}{y}, Q^2_{\gamma})
                             f_{\gamma/e}(y,Q^2_{WW})
\end{equation}
where 
\begin{equation}
df_{\gamma/e}(y,Q^2_{WW})=\frac{\alpha}{2\pi}[\frac{1+(1-x)^2}{xQ^2}
                          -2m^2_e \frac{x}{Q^4}]dQ^2
\end{equation}
$Q^2_{min} = m_e^2 x^2/(1-x)$ and $Q^2_{max}=Q^2_{WW}$ is defined by some
process scale that will be discussed later. After the substitution
of (5) and (7) into (4) and subtraction of counterterm (6) we finally get
\begin{eqnarray}
\sigma_{resolved}=\frac{\pi \sqrt{2}}{3} G_F m^2_W |V_{12}|^2
\int^1_{m_W^2/s} \int^1_x \frac{dx dy}{xy}
[f_{q_1 / {\gamma}}(\frac{x_1}{y},Q^2_{\gamma})-f^{LO}_{q_1 /
         {\gamma}}(\frac{x_1}{y},\frac{Q^2_{\gamma}}{\Lambda^2})]\\ \nonumber
            f_{\gamma/e}(y,Q^2_{WW}) f_{q_2/p}(\frac{m^2_W}{x_1s}, Q^2_p)
\end{eqnarray}
There are four scales $Q^2_{WW}, Q^2_{\gamma}, \Lambda^2, Q^2_p$ to be
defined in this formula. The Weizsacker-Williams scale $Q^2_{WW}$ can
be chosen equal to $\Lambda^2$. It is easy to justify this choice \cite{BVZ}
looking at the distribution $d\sigma/dlogQ^2_{q_1W}$ (Fig.2) calculated 
exactly for 
ten diagrams of $2 \rightarrow 4$ process in Fig.1. The flat part of
this distribution corresponds to the cross section behaviour $d\sigma/dt
\sim 1/t$ (not $1/t^2$ as it seems at the first sight, double poles
are cancelled),
and rapidly decreases starting from $\Lambda^2$. We always take proton
structure function scale $Q^2_p=m^2_W$. The values of $Q^2_{\gamma}, \Lambda^2$
are arbitrary and the final result for the sum of resolved and direct 
contributions (3) 
should not depend essentially from the choice of these two scales.

In Table 1 we show the results for resolved photon cross section
(9) calculated
by means of BASES MC integrator \cite{BASES}, using proton structure
functions MRS \cite{MRS} and CTEQ \cite{CTEQ},
photon structure functions DG \cite{DG}, LAC-G \cite{LAC} and GRV
\cite{GRV}.
Last two versions of gamma structure function are improved parametrizations
in the framework of the approach \cite{DG}.

One can see that if proton structure function is measured with rather
good accuracy and two parametrizations we are using give similar results
(consistent within the one standard deviation error of our MC integration),
photon structure function is still poorly known. LAC-G parametrization
gives the cross section regularly smaller than the values obtained
with the help of DG and GRV parametrizations.
We checked that our point 
obtained with the help of LAC2 and HMRS B structure functions at
$Q^2_{\gamma}=m_W^2/10$, $\Lambda=$ 5 GeV and equal to
$\sigma_{resolved}=$ 22 fb, is close to the value 24 fb given in \cite{BVZ}
at the same parameter values. So the possible correction to our result (9)
from the so-called 'finite terms' is around 8\%, which is much less than
the difference of results obtained by using different photon structure
functons. Negative cross section in Table 1 means that at a given scale
of $\Lambda$ strong double counting regime of direct and resolved
contrubutions takes place. In other words, most part of the resolved cross
section is already taken into account by the calculation for direct term
in (3).  

\begin{table}[t]
\begin{center}
\begin{tabular}{|c|c|c|c|c|c|c|c|c|}
\hline
\multicolumn{3}{|c|}{} & \multicolumn{3}{|c|}{MRS A} &
                                    \multicolumn{3}{|c|}{CTEQ3m} \\ \hline
$Q_{WW}$ &$Q_{\gamma}$ & $\Lambda$ & DG1 &LAC2 & GRV L0& DG1 & LAC2 & GRV L0
\\ \hline
0.2      & $ m_W$      &    0.2  &-11.6&-4.1  &-7.3  &-11.9& -4.2 & -7.6   \\
0.2      & $m_W/10 $   &    0.2  &-7.5 &2.0   &-3.3  &-7.6 & 2.7  & -3.1   \\
1.0      & $m_W$       &    1.0  &-5.6 &3.3   &-0.6  &-5.7 & 4.0  & -0.1   \\
1.0      & $m_W/10 $   &    1.0  &1.7  &13.4  & 6.5  &1.7  &14.8  & 7.4    \\
5.0      & $m_W$       &    5.0  &4.3  &14.7  & 9.8  &4.4  &16.3  & 11.0   \\
5.0      & $m_W/10 $   &    5.0  &15.7 &29.6  & 21.2 &16.0 &32.0  &23.1    \\
\hline
\end{tabular}
\end{center}
\caption{Resolved photon cross section of the process $e^- p \rightarrow
e^- \mu^+ \nu_{\mu} X$ in fb (see formula (9)). Different sets of
photon and proton structure functions were used.}
\end{table}

\newpage
  
\section{Complete tree level calculation and anomalous signal
            of $W$ in the distribitions}

\subsection{General framework}

Complete tree level calculation of 10 diagrams in Fig.1 with anomalous
three vector boson couplings (direct process)
and the following generation of particle distributions were done by means of 
CompHEP package \cite{CompHEP,WWgen}. The amplitude corresponding to 55
squared diagrams and interferences between diagrams was calculated
symbolically. In order to avoid $t$-channel poles, masses of electron
and quarks were kept nonzero.
\footnote{For this reason we need $\Lambda$ cut only for matching of
direct and resolved parts of cross section. For instance, if $m_u=$ 5 MeV,
$m_d=$ 10 MeV, $m_s=$ 0.2 GeV and $m_c=$ 1.3 GeV and there are no
kinematical cuts, $\sigma_{dir}(ep \rightarrow e \mu^+ \nu_{\mu} X)=$ 
102.5(5) fb in the standard case with MRS A structure functions.} 
Equivalent photon approximation was not
used. After that symbolic expressions are automatically converted to
FORTRAN codes and linked to special program for 
seven dimensional Monte-Carlo integration over four
particle phase space, and adaptive integration package VEGAS \cite{VEGAS}.
In the process of four particle phase space generation  
we introduce so-called kinematical regularization of the peaks \cite{IKP}
inherent to the amplitude under consideration. Especially these are 
$t$-channel gamma peaks in the $ee\gamma$ vertices of diagrams 1-4 (Fig.1), 
$\mu \mu \gamma$ vertex of diagram 1 and $W$-resonance peak in the 
diagrams with s-channel $W$-boson.

CompHEP package is a software product in the framework of one of a few
general approaches \cite{CompHEP,approaches} developed in recent time for the
analysis of multiparticle exclusive states at new colliders,
when hundreds of Feynman diagrams contribute to the amplitude 
and should be exactly calculated. More details can be found in \cite{WWgen}.

In our calculations we used the Breit-Wigner propagator with constant width
for the $W$-boson.
Generally speaking, if we have some complete tree level set of diagrams,
straightforward replacement
of lowest order vector boson propagator by the propagator with finite width
violates
gauge invariance of the amplitude and can break gauge cancellations
between diagrams, leading to numerically unstable false results. For this
reason we used the well-known "overall" form of
propagator replacement \cite{BVZ} when the entire amplitude is multiplied
by a factor
\begin{equation}
\frac{p^2_W-m^2_W}{p^2_W-m^2_W+i m_W \Gamma_W}
\end{equation}
Generally speaking in other cases this prescription could affect strongly
nonresonant
terms in the amplitude \cite{BDD} (in the region of phase space where
$p^2 \sim m^2_W$), but the case under consideration is free from this
difficulty.

\begin{table}[h]
\begin{center}
\begin{tabular}{|c|c|c|c|c|c|c|}
\hline
\multicolumn{7}{|c|}{$\Lambda=$ 0.2 GeV} \\ \hline
  $\lambda$ & $k$ & $eu\rightarrow e \mu^+ \nu_{\mu} d$
                  & $e \bar d \rightarrow e \mu^+ \nu_{\mu} \bar u$  
                  & $eu\rightarrow e \mu^+ \nu_{\mu} s$
                  & $e \bar s\rightarrow e \mu^+ \nu_{\mu} \bar c$
                  & $\sigma_{tot}$ \\ \hline
      0  &  1  &  68.2(4)& 15.2(1) & 3.5(0)  &5.2(0) &92.1(5) \\
      1  &  1  &  72.6(5)& 15.6(1) & 3.7(0)  &5.3(0) &97.2(6)  \\
      0  &  0  &  52.6(5)& 13.6(1) & 2.7(0)  &4.6(0) &73.5(6)  \\
      0  &  2  &  99.6(6)& 17.2(2) & 5.1(0)  &6.3(0) &128.2(8) \\ \hline
\multicolumn{7}{|c|}{$\Lambda=$ 1.0 GeV} \\ \hline
      0   &  1  &  61.2(4) &12.4(2) & 3.1(0) &5.2(0) &81.9(6) \\   
      1   &  1  &  66.0(4) &12.6(1) & 3.4(0) &5.3(0) &87.3(5) \\
      0   &  0  &  45.3(3) &10.8(1) & 2.3(0) &4.6(0) &63.0(4) \\   
      0   &  2  &  92.7(6) &14.9(1) & 4.8(0) &6.3(0) &118.7(7)\\ \hline
\multicolumn{7}{|c|}{$\Lambda=$ 5.0 GeV} \\ \hline
      0   &  1  &  52.5(3) &8.9(0) & 2.7(0) &4.5(0) &68.6(3) \\
      1   &  1  &  57.4(4) &9.0(0) & 2.9(0) &4.6(0) &73.9(4) \\
      0   &  0  &  36.4(3) &7.4(0) & 1.9(0) &3.9(0) &49.6(3) \\
      0   &  2  &  83.8(5) &11.1(0) & 4.3(0) &5.6(0)&104.8(5) \\ \hline
\end{tabular}
\end{center}
\caption{ Total cross sections (fb) of the main partonic $W^+$
          producing processes (see 10
          diagrams in Fig.1) in the reaction $e^- p \rightarrow
          e^- \mu^+ \nu_{\mu} X$ for the various sets of $k$,
          $\lambda$.
           Invariant kinematical cut $\protect{\Lambda^2=(p_q-p_{W})^2}$.
          Proton
          structure function MRS A. One standard deviation error
          of Monte-Carlo integration for the last digit
          is indicated in brackets.}
\end{table}

\begin{table}[h]
\begin{center}
\begin{tabular}{|c|c|c|c|c|c|c|}
\hline
\multicolumn{7}{|c|}{$\Lambda=$ 0.2 GeV, $E_{\mu}>$ 10 GeV, missing
$p_T>$ 20 GeV} \\ \hline
  $\lambda$ & $k$ & $eu\rightarrow e \mu^+ \nu_{\mu} d$
                  & $e \bar d \rightarrow e \mu^+ \nu_{\mu} \bar u$
                  & $eu\rightarrow e \mu^+ \nu_{\mu} s$
                  & $e \bar s\rightarrow e \mu^+ \nu_{\mu} \bar c$
                  & $\sigma_{tot}$ \\ \hline
      0  &  1  &  49.2(4)& 11.1(1) & 2.5(0)  &3.6(0) &66.4(5) \\
      1  &  1  &  54.7(5)& 11.4(1) & 2.8(0)  &3.7(0) &72.6(6)  \\
      0  &  0  &  36.6(4)&  9.8(1) & 1.8(0)  &3.0(0) &51.2(5)  \\
      0  &  2  &  75.9(5)& 12.8(2) & 3.9(0)  &4.5(0) &97.1(7) \\ \hline
\end{tabular}
\end{center}
\caption{ The same as in Table 2 with the kinematical cuts imposed
           $E_{\mu}>$ 10 GeV, missing $p_T>$ 20 GeV. 
          These cuts are
          used to exclude the misidentification backgounds in
          the electron channel (see section 4.2). }
\end{table}

The accuracy of our Monte Carlo calculation of $\sigma_{dir}$ is 
usually 0.6-0.8\% (see Table 2).
This choice is related to the precision of proton structure function
parametrization. We checked that if we replace the proton structure 
set MRS A \cite{MRS} that we are using by the set CTEQ3m 
\cite{CTEQ}, the relative difference of results obtained with these 
two sets does not exceed 1\%.
  
Four partonic processes from 12 possible give the main contribution to
$W$ boson production. We present the results of total cross section 
calculation for the main subprocesses of the $W^+$ production channel 
$e^- p \rightarrow e^- \mu^+ \nu_{\mu} X$
in Table 2. Main contribution comes from the subprocess
$e^- u \rightarrow e^- \mu^+ \nu_{\mu} d$.
The remaining 8 partonic
reactions have very small individual total cross sections and the 
sum of their contributions is of order 1 $fb$.
The cross section of
$W^-$ production channel $e^- p \rightarrow e^- \mu^- \bar \nu_{\mu} X$
is compared with the case of $W^+$ production in Table 4. In the case
of $W^-$ production the main partonic subprocess is
$e^- d \rightarrow e^- \mu^+ \nu_{\mu} u$
and the total rate is slightly smaller because there are less
$d$-quarks in the proton than $u$-quarks. Resolved photon contribution
from Table 1 is also indicated. One can see that indeed rather weak
dependence of $\sigma_{dir}+\sigma_{res}$ from the cutoff $\Lambda$
takes place in so far as the decrease of direct part with the growth
of $\Lambda$ is compensated by the increase of resolved part.

We already mentioned in Section 3 that $Q^2_p=m^2_W$ was always
taken as the momentum transferred scale in the proton structure
functions. It is important to find out how the changes of $Q^2_p$
affect the total cross section value and compare the possible deviation
of total rate
caused by the change of hadronic scale with the deviation of total
rate coming from anomalous three vector boson couplings. The
uncertainties coming from the value of $Q^2_p$ should be less than
the effect of anomalous couplings to make the phenomenological
restrictions based on the value of total rate more meaningful.
We show the numbers for total cross section of the main $W^+$ producing
partonic process $eu \rightarrow e \mu \nu_{\mu} d$ calculated for
proton structure function scales $m_W^2/2$ and $2m_W^2$ in Table 5.
These values are taken as illustrative ones because in real
partonic processes the contributions from so small/large values of $Q^2_p$ 
are negligible. We can see that 
the total cross section deviation from the standard choice of
hadronic scale $Q^2_p=m_W^2$ is around 2.5 - 3\%, while the effects
of anomalous $W$ couplings are much larger. For instance, the effect 
coming from anomalous $k$-term in 20 - 30\% and the effect of anomalous
$\lambda$ term is 6 - 8\% (Table 2).

\begin{table}[ht]
\begin{center}
\begin{tabular}{|c|c|c|c|c|c|} \hline
\multicolumn{3}{|c|}{}& \multicolumn{3}{|c|}{$\Lambda$} \\ \hline
                      &$\lambda$& $k$ &  0.2  &  1.0  &   5.0 \\ \hline
                      &    0    &  1  &92.1   &81.9   &68.6  \\
$\sigma_{dir}(W^+)$   &    1    &  1  &97.2   &87.3   &73.9  \\
                      &    0    &  0  &73.5   &63.0   &49.6  \\
                      &    0    &  2  &128.2  &118.7  &104.8  \\ \hline
$\sigma_{res}(W^+)$   &\multicolumn{2}{|c|}{}
                                      &-4.1   &3.3    &14.7   \\ \hline
$\sigma_{dir}+\sigma_{res}$, $W^+$, SM& \multicolumn{2}{|c|}{}
                                      &88.0   &85.2   &83.3     \\ \hline
\hline
                      &    0    &  1  &80.3   &68.6   &52.4  \\
$\sigma_{dir}(W^-)$   &    1    &  1  &82.7   &70.6   &53.9   \\
                      &    0    &  0  &69.5   &57.6   &41.4   \\
                      &    0    &  2  &97.6   &86.0   &69.5   \\ \hline
$\sigma_{res}(W^-)$   &\multicolumn{2}{|c|}{}
                                      &-11.0  &-3.8   &7.9     \\ \hline
$\sigma_{dir}+\sigma_{res}$, $W^-$, SM& \multicolumn{2}{|c|}{}
                                      &69.3   &64.8   &60.3    \\ \hline
\hline
\end{tabular}
\end{center} 
\caption{Total cross sections (fb) of the reactions
$e^- p \rightarrow e^- \mu^+ \nu_{\mu} X$ ($W^+$ production) and
$e^- p \rightarrow e^- \mu^- \bar \nu_{\mu} X$ ($W^-$ production)
in the case of standard (SM) and anomalous three vector boson interaction
at different values of $\Lambda$ cutoff parameter. Proton structure
function MRS A. In the calculation of resolved photon contribution
gamma structure function LAC-G2 was taken at the scale $m_W$ .}
\end{table}

\subsection{The reactions $e^- p \rightarrow e^- e^+ \nu_e X$, 
               $e^- p \rightarrow e^- e^- \bar \nu_e X$ and
 the comparison with EPVEC generator}

Total number of diagrams in the channels $e^- p \rightarrow e^- e^+ \nu_e X$
and $e^- p \rightarrow e^- e^- \bar \nu_e X$ is 20 (ten additional
diagrams to Fig.1 with $t$-channel $W$-boson exchanges appear). However at 
HERA energy the correction coming from these additional 
diagrams is very small because they do not contain $t$-channel photon
poles. Usually this weak contribution is neglected.
Such approximation was used in the recent simulation for HERA \cite{noyes}
(by means of EPVEC generator \cite{BVZ}), where event topology
and realistic kinematical cuts were considered in more details.

The electron channels $e^- e^+ \nu_e X$ and $e^- e^- \bar \nu_e X$
are more difficult for experimental study than the muon channels
$e^- \mu^+ \nu_{\mu} X$ and $e^- \mu^- \bar \nu_{\mu} X$. First,
large background from the neutral current deep inelastic scattering process
$ep \rightarrow eX$ appears in the case when the final jet energy is not 
completely registered in the hadronic calorimeter and for this reason some 
missing
$p_T$ is observed. Second, large background from the charged current
DIS process $ep \rightarrow \nu X$ appears in the case when $\pi^0$
from the final jet is misidentified as $e^-(e^+)$ giving again the
final state with $e^-(e^+)$, jet and missing $p_T$. In order to
suppress these misidentification backgrounds the following
kinematical cuts were used in \cite{noyes}: (1) isolated electron
with the energy $E_e >$ 10 GeV (2) missing transverse momentum
$p_T >$ 20 GeV are required. The requirement of isolated electromagnetic
cluster is removed for the case of muonic channels, and 10 GeV energy
cut seems too strong for the final muon, since good reconstruction of several
GeV muons is available in ZEUS and H1 detectors. Missing
transverse momentum cut also seems not necessary for the muon channels
(scattered lepton is different from $W$ decay lepton). 
 
We checked by direct calculation that in the absence of missing
$p_T$ cut, muon energy cut at 2-3 GeV practically does not affect
the numbers for total cross sections shown in Table 2 and Table 4.
The combination of muon energy cut at 10 GeV and missing $p_T$ cut
at 20 GeV gives the cross sections by 30-35\% smaller than before
cuts (see Table 3). Following \cite{noyes} we shall consider electron 
channels in the approximation of 10 diagrams subset (Fig.1) with the
kinematical cuts $E_e >$ 10 GeV and missing $p_T >$ 20 GeV.

In Fig.3 we show the distributions of energy, angle with the 
beam and transverse momentum for electron, muon and final quark in the
muonic channel $e^- \mu^+ \nu_{\mu} X$, obtained by means of
CompHEP (Standard Model case). The same distributions for electron 
channel, obtained with the help of EPVEC generator, can be found in 
\cite{noyes}.
(We are using in (3) $\Lambda=$ 0.2 GeV, while the distributions 
in \cite{noyes} are calculated using $\Lambda=$ 5 GeV.)
Although the normalization is not indicated in \cite{noyes}, one
can observe that the agreement of the shapes is satisfactory. 
Soft muons in the distributions $d\sigma/dE_{\mu}$ and $d\sigma/dp_{T \mu}$
come from the ladder diagrams 1,6,7 in Fig.1. Jets at
the angle 180 degrees with the proton beam appear from diagram 3,
Fig.1, when the quasireal photon produces two quarks collinear to
initial electron. Soft muons and backward jets are absent in
\cite{noyes}, because besides the $E_e$ and missing $p_T$ cuts mentioned 
above, the EPVEC generator contains build-in
cuts \cite{prcom} separating some region of phase space near the $W$ pole.

\subsection{Sensitivity to anomalous couplings}

Let us return to our calculation with anomalous interaction of
vector bosons.
We can estimate approximately the possibilities of HERA in
the detection of anomalous couplings using a simple criteria
(see, for instance, \cite{SS}) for the number of events $N$
that is necessary to observe $\Delta \sigma$ deviation from the total cross 
section value $\sigma$: 
\begin{equation}
\frac{\Delta \sigma}{\sigma} \sim \frac{1}{\sqrt{N}}
\end{equation}
It follows from Tables 2,4 that at the integrated luminosity of HERA 
$L=$ 200 $pb^{-1}$ in the channels of $W^+$ and $W^-$
production $e^- \mu^+ \nu_{\mu} q$ and $e^- \mu^- \bar \nu_{\mu} q$
we shall have about 35 events/year. Deviation
of $\lambda$ in the vertex (1) from the zero standard value 
$\Delta \lambda =1$
gives us 5\% deviation in the total $W$ production rate and we need
about 400 events to observe it. However the deviation $\Delta k$=1 
($\Delta k = k-1$) changes the total cross section by 20-40 \% and 
less than 25 events will be needed for some experimental evidence.
$W^+$ and $W^-$ production in the channels
$e^- e^+ \nu_{e} q$ and $e^- e^- \bar \nu_{e} q$ will give slightly
less reliably reconstructed events than in muonic channels (kinematical
cuts must be introduced to tune off the misidentification backgrounds), 
so in total around 60 $W$ bosons/year decaying to electrons and muons could
be observed. It follows that it will be difficult to improve
CDF and D0 limits on $\lambda$ \cite{Tevatron}, but some
improvenent of $\Delta k$ restriction could be possible. 

More precisely, systematic errors on the detector acceptance $A$
and the uncertainty in the luminosity measurement $L$ should be taken 
into account. The former are estimated on the level of 2\%(1\%) for
the integrated luminosity of order $10^2$ $pb^{-1}$($10^3$ $pb^{-1}$) and 
the latter is taken to be 2\%. The uncertainty of the total cross section
measurement has the form \cite{noyes}
\begin{equation}
\frac{\Delta \sigma}{\sigma}=(\frac{1}{N}+
                               (\frac{\Delta L}{L})^2+
                               (\frac{\Delta A}{A})^2)^{\frac{1}{2}}
\end{equation}
The acceptances in both electron and muon channels are taken to be 65\%.
From equation (12) we derive the following limits for  
$\Delta k$ and $\lambda$, giving the observable deviation of
total cross section from the Standard Model value at 68\% and 95\%
confidence level:
\begin{eqnarray*}
-1.70 < \lambda < 1.70,\phantom{xxxx} -1.05 < \Delta k < 0.48,
                       \phantom{xxxx}  68\% \quad CL \\
-2.24 < \lambda < 2.24,\phantom{xxxxxxxxxxxx} \Delta k < 0.89,
                       \phantom{xxxx}  95\% \quad CL 
\end{eqnarray*}
at the integrated luminosity 200 $pb^{-1}$, and
\begin{eqnarray*}
-1.03 < \lambda < 1.03,\phantom{xxxx} -0.31  <\Delta k < 0.27,
                       \phantom{xxxx}  68\% \quad CL \\
-1.75 < \lambda < 1.75,\phantom{xxxx} -0.58  <\Delta k < 0.46,
                       \phantom{xxxx}  95\% \quad CL   
\end{eqnarray*}
at the integrated luminosity 1000 $pb^{-1}$. Here only one coupling
from the pair ($\lambda$, $k$) is assumed to be different from
the SM value. We do not indicate 
the negative 95\%$CL$ $\Delta k$ limit at the integrated luminosity 200 
$pb^{-1}$ because the cross section deviation from the SM value stops to
increase starting from $\Delta k \sim$ -1.5 and the effect
cannot be observed with small statistics. However, positive  
$\Delta k$ limit at the same luminosity
is competitive with the early expectations from LEP2 \cite{LEP2}.         
These limits could be of course improved by taking into account the
channels $W \rightarrow jets$ with final electron and 
three jets in the final state. Low acceptance in the jets channel 
($\sim$ 20\%, \cite{noyes}) and complicated situation with the 
separation of QCD backgrounds requires an independent careful study, 
and we are not considering this possibility here. 
 
\begin{table}[hb]
\begin{center}
\begin{tabular}{|c|c|c|c|c|}
\hline
\multicolumn{2}{|c|}{} & $m_W^2/2$  & $m_W^2$ & $2 m_W^2$ \\ \hline
$\lambda$ & $k$   &\multicolumn{3}{|c|}{}                \\ \hline
    0     &      1     & 70.6(5)    & 68.2(4) & 66.3(6)  \\  
    1     &      1     & 75.5(5)    & 72.6(5) & 71.1(4)  \\
    0     &      0     & 54.1(4)    & 52.6(5) & 51.2(4)  \\
    0     &      2     &102.3(6)    & 99.6(6) & 96.2(6)  \\ \hline
\end{tabular}
\end{center}
\caption{ Total cross section (fb) of the main $W^+$ producing partonic
process $eu \rightarrow e \mu \nu_{\mu} d$ calculated using three $Q^2$
scales in the parametrizations of proton structure function MRS A.}
\end{table}

Of course the calculation of total rate is very important and the
ratio $\sigma_{tot}(W)/\sigma_{tot}(Z)$ that was considered in
\cite{KLS} could be the clear indicator to anomalous gauge boson
coupling. However the only way to see definetly if the deviation
of the ratio is really due to anomalous $WW\gamma$ interaction but
not caused by some other reason, is to inspect what regions of 
phase space are affected by anomalous $W$ interaction dynamics and 
how they are affected. It is natural to use, as proposed in \cite{BZ},
the distributions of final quark and muon transverse momenta.

We show the distributions of final quark jet transverse momentum in
Fig.4,5. It follows from symbolic calculation in $2 \rightarrow 2$ 
approximation that the cross section depends from $\lambda$ quadratically
\cite{KLS}. We checked at complete tree level $2 \rightarrow 4$ that
this is true for $\sigma_{tot}$ at about 1\% accuracy and no 
difference at positive and negative values of $\lambda$ is observed 
in the distributions. The deviation of $k$ is clearly seen 
(Fig.5), $d\sigma/dp_{T}$ becomes harder when $k$ is less than standard
value $k=1$.  

The distributions of final muon transverse momentum are shown in Fig.6,7. 
Similar to the previous case of jet $p_T$ distribution, the effect
coming from $\lambda$ is very small and the dependence of distribution 
from $k$ is rather strong. The forward
and backward slopes of $W$ peak can be slightly shifted if we
take into account $W$ production by the resolved photon (see the details
in\cite{BVZ}), but this shift is the same for standard and nonstandard 
cases.

The distributions of final muon rapidity for the standard and
anomalous cases are shown in Fig.8. The
direction of proton beam was chosen as the direction of positive
rapidity axis.

\begin{table}[h]
\begin{center}
\begin{tabular}{|c|c|c|c|c|} \hline
 $\lambda$ & $k$ & exact result& $Q_{WWA}=m_W$ & $Q_{WWA}$=20 GeV \\ \hline
     0     &  1  & 61.2(4)& 66.6(8)      &  61.1(5)         \\
     1     &  1  & 66.0(4)& 70.9(6)      &  64.5(5)         \\
     0     &  0  & 45.3(3)& 49.4(5)      &  46.5(4)        \\ 
     0     &  2  & 92.7(6)& 96.7(9)      &  88.1(8)          \\ \hline  
\end{tabular}
\end{center} 
\caption{Comparison of exact calculation (fb) for the process
$e^- u \rightarrow e^- \mu^+ \nu_{\mu} d$ with equivalent photon
approximation for the process $\gamma u \rightarrow \mu^+ \nu_{\mu} d$
calculated at the scale $Q_{WW}$ (8) equal to $m_W$ and 20 GeV,  
$\Lambda=$1.0 GeV.}
\end{table}

An important point concerning the $p_T$ distributions of the quark
and muon for the direct process is their sensitivity to higher order 
QCD corrections, which could be large. As it was stated in \cite{KLS,BZ},
the integration close to $u$-channel pole in the diagram 3, Fig.1 involves
the momenta of the order $\Lambda_{QCD}$ in the small $p_T$ region,
when the QCD corrections can be expected to be significant. Total
cross section in the absence of $u$-channel or $p_T$ cuts contains
some degree of uncertainty. It was mentioned in \cite{BZ} that in
connection with normalization uncertainty of the cross section the
events when the jet escapes detection and only the lepton with missing
$p_T$ are observed could provide an important test on the normalisation
of $W$ production rate.  

Finally we would like to discuss the question of equivalent photon
approximation \cite{WWA} accuracy in our case. In Table 6 we compare
exact result for the partonic subprocess $e u \rightarrow e \mu^+
\nu_{\mu} d$ and the equivalent gamma approximation calculation 
for the process $\gamma u \rightarrow \mu^+ \nu_{\mu} d$. In the latter case
intial photon momentum is distributed according to (8) where we used
two different choices of $Q_{WW}$. One can see that the 'natural'
choice $Q_{WW}=m_W$ overestimates the cross section by 8\% (let
us remind (see Table 4) that the effect of $\Delta \lambda=$ 1 
is 5\%). It is possible of course to adjust $Q_{WW}$ which is not
strictly fixed at any value, but defined by some typical process dependent
momentum transferred scale, and get agreement 
of exact and WW cross sections for the standard case $\lambda=0$, $k=1$; 
it turns out that the
corresponding value is $Q_{WW}=$ 20 GeV. However, after fixing of this value
in the anomalous case $\lambda=0$, $k=2$ again we observe 5\% deviation.
Equivalent approximaions become too rough if precise separation of the
signal is needed.

\section{Conclusion}

We presented the results of complete tree level calculation
for the $W$ boson production processes at the energy of HERA collider,
introducing anomalous C and P conserving three vector boson couplings (1). 
The main $W^+$ and $W^-$ production channels 
$e^- p \rightarrow e^- l^+ \nu_{l} X$ and 
$e^- p \rightarrow e^- l^- \bar \nu_{l} X$ ($l=e, \, \mu$)were considered. 
Following the earlier publications
\cite{BVZ,KLS} we separated the phase space at some scale of
momentum transferred from the constituent quark to $W$-boson
in order to take into account the resolved photon contribution 
to the total rate.
Resolved photon part was calculated in the structure function approach, 
using new parametrizations of photon and proton distribution functions.
Perturbative (direct) part of the cross section was considered
by means of CompHEP package \cite{CompHEP},\cite{WWgen}, when the
tree level $2 \rightarrow 4$ 
amplitude, corresponding to ten diagrams for each of $W^+$ and $W^-$
production processes, is calculated exactly without any approximations. 
Some uncertainty in the normalization of total cross section
exists for the reason of possibly large QCD corrections in the
phase space region near the $u$-channel quark pole.
In the muonic channels under consideration the total
cross section is equal approximately to 150-160 fb, giving about
35 events/year at the integrated luminosity 200 $pb^{-1}$.
Kinematical cuts are necessary in the electron channels for 
separation of misidentification backgrounds, and the number of
identifiable events from $W \rightarrow e \nu_e$ is slightly smaller.

We show explicitly what regions of phase space are affected by
anomalous three vector boson interaction dynamics. In particular
it follows from our analysis that even at the integrated luminosity
1000 $pb^{-1}$ it will be extremely difficult
to separate anomalous $\lambda$ term effect in (1) when $\Delta \lambda$
is less than 1.5 (as already restricted by Tevatron data), but rather 
easy to observe anomalous $k$ term effect, when $\Delta k$ is of order
0.4-0.8, which is strongly competitive with LEP2 possibilities. 

\begin{center}
{\bf Acknowledgements}
\end{center}
The authors are grateful to Edward Boos, C.S.Kim and 
Alexander Pukhov for useful discussions. M.D.
would like also to thank the Center for Theoretical Physics, Seoul
National University, for hospitality.
The research of M.D. was partially supported by INTAS grant 93-1180ext 
and RFBR grant 96-02-19773a.

\newpage

.

\begin{center}
{\bf \large Figures}\\
\end{center}

.

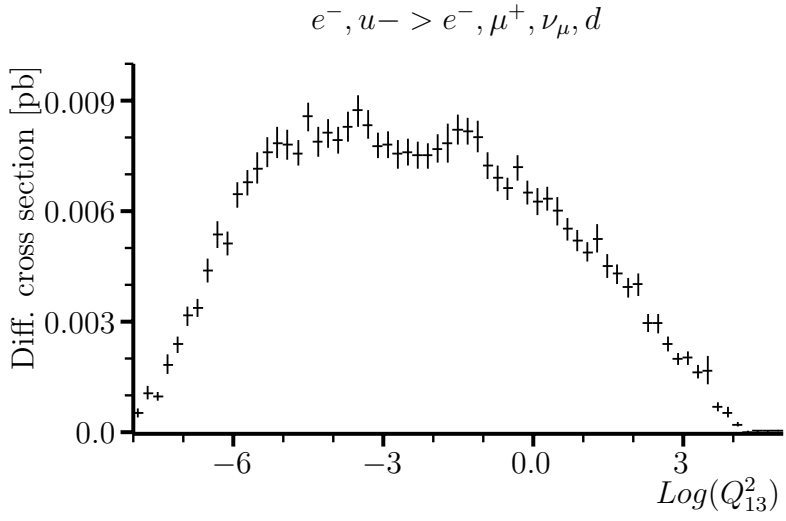
\begin{figure}[h]
{\def\chepscale{1.0} 
\unitlength=\chepscale pt
\SetWidth{0.7}      
\SetScale{\chepscale}
\normalsize    
\begin{picture}(300,200)(0,0)
\Text(168.2,199.4)[t]{$e^- , u     ->  e^- , \mu^+ , \nu_{\mu} , d$}
\LinAxis(45.90,36.72)(290.82,36.72)(4.333,3,-4,1.000,1.5)
\Text(83.4,29.9)[t]{$-6$}
\Text(140.2,29.9)[t]{$-3$}
\Text(196.5,29.9)[t]{$0.0$}
\Text(253.3,29.9)[t]{$3$}
\Text(290.8,20.3)[rt]{$Log(Q^2_{13})$}
\LinAxis(45.90,36.72)(45.90,176.27)(3.348,3,4,-0.044,1.5)
\Text(39.2,37.3)[r]{$0.0$}
\Text(39.2,79.1)[r]{$0.003$}
\Text(39.2,120.9)[r]{$0.006$}
\Text(39.2,162.1)[r]{$0.009$}
\rText(11.3,176.3)[tr][l]{Diff. cross section [pb]}
\Line(288.7,37.9)(288.7,37.9) 
\Line(286.6,37.9)(290.8,37.9) 
\Line(285.0,37.9)(285.0,37.3) 
\Line(282.9,37.9)(286.6,37.9) 
\Line(281.2,37.9)(281.2,37.3) 
\Line(279.1,37.9)(282.9,37.9) 
\Line(277.5,37.9)(277.5,36.7) 
\Line(275.4,37.3)(279.1,37.3) 
\Line(273.7,41.2)(273.7,39.5) 
\Line(271.6,40.1)(275.4,40.1) 
\Line(270.0,46.9)(270.0,42.9) 
\Line(267.9,44.6)(271.6,44.6) 
\Line(266.2,48.6)(266.2,45.2) 
\Line(264.1,46.9)(267.9,46.9) 
\Line(262.4,66.1)(262.4,55.4) 
\Line(260.4,60.5)(264.1,60.5) 
\Line(258.7,62.7)(258.7,57.6) 
\Line(256.6,59.9)(260.4,59.9) 
\Line(254.9,67.8)(254.9,62.7) 
\Line(252.9,65.5)(256.6,65.5) 
\Line(251.2,67.2)(251.2,62.7) 
\Line(249.1,65.0)(252.9,65.0) 
\Line(247.4,73.4)(247.4,67.8) 
\Line(245.3,70.6)(249.1,70.6) 
\Line(243.7,81.9)(243.7,74.6) 
\Line(241.6,78.5)(245.3,78.5) 
\Line(239.9,81.9)(239.9,75.1) 
\Line(237.8,78.5)(241.6,78.5) 
\Line(236.2,97.2)(236.2,88.7) 
\Line(234.1,93.2)(237.8,93.2) 
\Line(232.4,95.5)(232.4,88.1) 
\Line(230.3,92.1)(234.1,92.1) 
\Line(228.2,100.6)(228.2,93.2) 
\Line(226.6,97.2)(230.3,97.2) 
\Line(224.5,104.5)(224.5,95.5) 
\Line(222.8,100.0)(226.6,100.0) 
\Line(220.7,115.8)(220.7,105.1) 
\Line(219.1,110.2)(222.8,110.2) 
\Line(217.0,109.0)(217.0,101.7) 
\Line(215.3,105.1)(219.1,105.1) 
\Line(213.2,113.6)(213.2,105.6) 
\Line(211.5,109.6)(215.3,109.6) 
\Line(209.5,118.1)(209.5,109.6) 
\Line(207.8,114.1)(211.5,114.1) 
\Line(205.7,126.0)(205.7,115.3) 
\Line(204.0,120.9)(207.8,120.9) 
\Line(201.9,129.9)(201.9,120.9) 
\Line(200.3,125.4)(204.0,125.4) 
\Line(198.2,129.4)(198.2,119.2) 
\Line(196.5,124.3)(200.3,124.3) 
\Line(194.4,132.2)(194.4,123.2) 
\Line(192.8,127.7)(196.5,127.7) 
\Line(190.7,141.8)(190.7,132.2) 
\Line(189.0,137.3)(192.8,137.3) 
\Line(186.9,133.3)(186.9,124.9) 
\Line(185.3,129.4)(189.0,129.4) 
\Line(183.2,137.9)(183.2,128.2) 
\Line(181.5,133.3)(185.3,133.3) 
\Line(179.4,142.9)(179.4,132.8) 
\Line(177.7,137.9)(181.5,137.9) 
\Line(175.7,154.8)(175.7,142.9) 
\Line(174.0,148.6)(177.7,148.6) 
\Line(171.9,155.9)(171.9,145.8) 
\Line(170.2,150.8)(174.0,150.8) 
\Line(168.2,157.1)(168.2,145.8) 
\Line(166.1,151.4)(170.2,151.4) 
\Line(164.4,153.7)(164.4,139.0) 
\Line(162.3,146.3)(166.1,146.3) 
\Line(160.6,149.7)(160.6,139.0) 
\Line(158.6,144.1)(162.3,144.1) 
\Line(156.9,146.3)(156.9,136.7) 
\Line(154.8,141.8)(158.6,141.8) 
\Line(153.1,146.9)(153.1,136.7) 
\Line(151.0,141.8)(154.8,141.8) 
\Line(149.4,148.0)(149.4,137.9) 
\Line(147.3,142.9)(151.0,142.9) 
\Line(145.6,147.5)(145.6,136.7) 
\Line(143.5,142.4)(147.3,142.4) 
\Line(141.9,150.8)(141.9,140.7) 
\Line(139.8,145.8)(143.5,145.8) 
\Line(138.1,150.3)(138.1,140.7) 
\Line(136.0,145.2)(139.8,145.2) 
\Line(134.4,158.8)(134.4,148.0) 
\Line(132.3,153.1)(136.0,153.1) 
\Line(130.6,164.4)(130.6,152.5) 
\Line(128.5,158.8)(132.3,158.8) 
\Line(126.8,158.2)(126.8,146.9) 
\Line(124.8,152.5)(128.5,152.5) 
\Line(123.1,152.5)(123.1,142.4) 
\Line(121.0,147.5)(124.8,147.5) 
\Line(119.3,155.4)(119.3,144.6) 
\Line(117.2,150.3)(121.0,150.3) 
\Line(115.6,152.5)(115.6,141.2) 
\Line(113.5,146.9)(117.2,146.9) 
\Line(111.8,161.6)(111.8,150.8) 
\Line(109.7,156.5)(113.5,156.5) 
\Line(108.1,147.5)(108.1,137.9) 
\Line(106.0,142.4)(109.7,142.4) 
\Line(103.9,151.4)(103.9,140.1) 
\Line(102.2,145.8)(106.0,145.8) 
\Line(100.1,152.5)(100.1,140.7) 
\Line(98.5,146.3)(102.2,146.3) 
\Line(96.4,148.6)(96.4,137.3) 
\Line(94.7,142.9)(98.5,142.9) 
\Line(92.6,142.9)(92.6,131.1) 
\Line(91.0,136.7)(94.7,136.7) 
\Line(88.9,136.2)(88.9,126.6) 
\Line(87.2,131.6)(91.0,131.6) 
\Line(85.1,131.6)(85.1,122.0) 
\Line(83.4,127.1)(87.2,127.1) 
\Line(81.4,113.0)(81.4,104.0) 
\Line(79.7,108.5)(83.4,108.5) 
\Line(77.6,116.9)(77.6,106.8) 
\Line(75.9,111.9)(79.7,111.9) 
\Line(73.9,102.8)(73.9,93.8) 
\Line(72.2,98.3)(75.9,98.3) 
\Line(70.1,87.6)(70.1,80.8) 
\Line(68.4,84.2)(72.2,84.2) 
\Line(66.3,84.7)(66.3,77.4) 
\Line(64.7,81.4)(68.4,81.4) 
\Line(62.6,73.4)(62.6,67.2) 
\Line(60.9,70.6)(64.7,70.6) 
\Line(58.8,66.7)(58.8,59.3) 
\Line(57.2,62.7)(60.9,62.7) 
\Line(55.1,52.5)(55.1,49.2) 
\Line(53.4,50.8)(57.2,50.8) 
\Line(51.3,54.8)(51.3,49.7) 
\Line(49.7,52.0)(53.4,52.0) 
\Line(47.6,46.3)(47.6,42.9) 
\Line(45.9,44.6)(49.7,44.6) 
\end{picture}
}
\caption{Distribution over logarithm of momentum transferred squared
$Q^2_{13}=t_{13}=(p_e^{in}-p_e^{out})^2$. Flat part of
the distribution corresponds to the behaviour $d\sigma/dt \sim 1/t_{13}$}
\end{figure}

\newpage

\unitlength 1.0cm

\begin{figure}
\begin{picture}(17,17)
\put(-1,0){\epsfxsize=17cm
         \epsfysize=17 cm \leavevmode \epsfbox{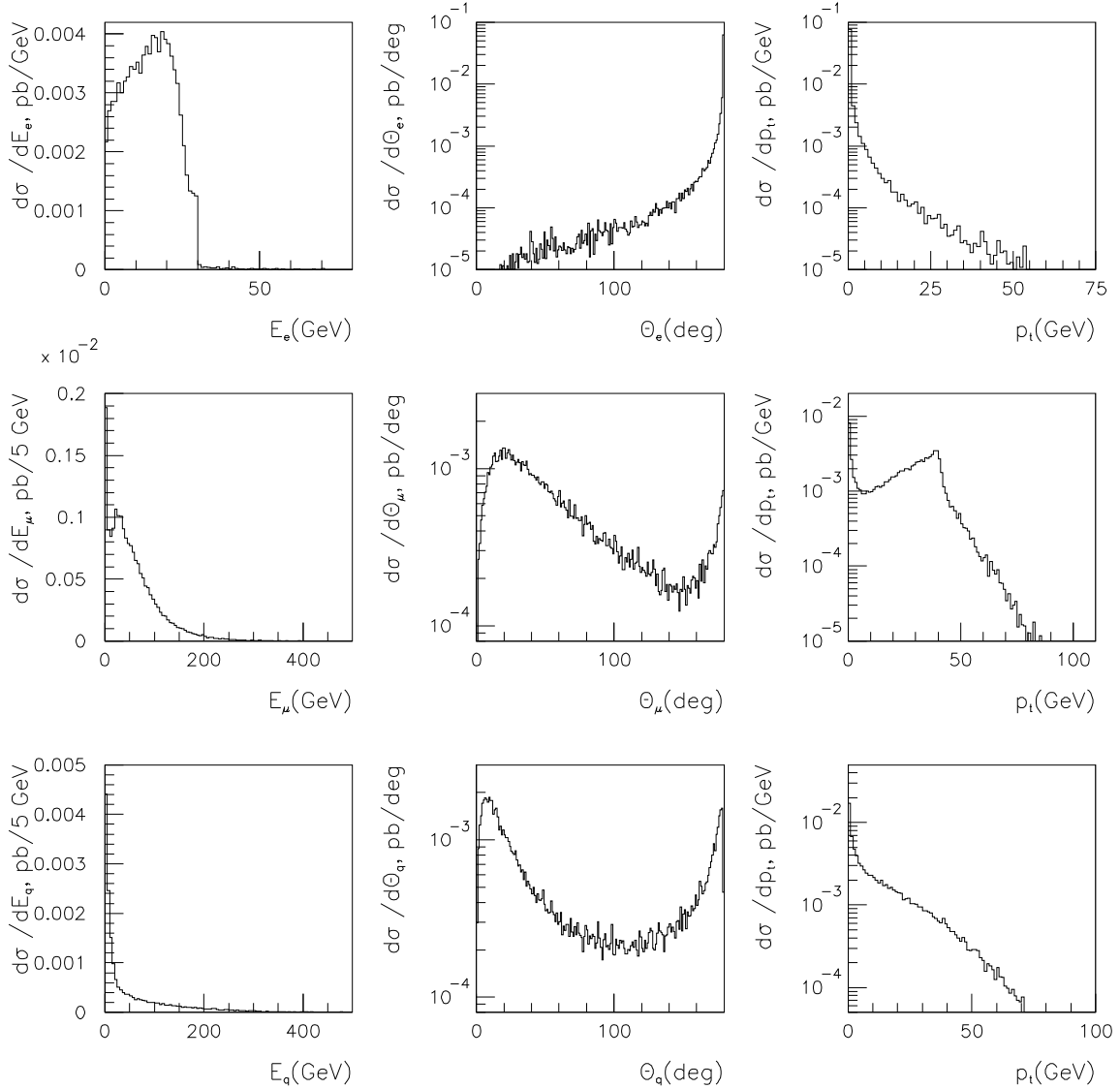}}
\end{picture}
\caption{First row of plots - distributions of the electron energy,
scattering angle and transverse momentum in the process
$e^- p \rightarrow e^- \mu^+ \nu_{\mu} X$. Second row of plots -
distributions of the muon energy, muon scattering angle and transverse
momentum. Third row of plots - distributions of the quark energy, angle
and transverse momentum for the same process. No kinematical cuts, all  
calculations were done by means of CompHEP [18]
($\Lambda=$ 0.2 GeV) to be compared with
the same distributions obtained by means of EPVEC generator
($\Lambda=$ 5.0 GeV), see
[7]}
\end{figure}

\newpage

\begin{figure}
\begin{picture}(15,15)
\put(-6,-3){\epsfxsize=19cm
         \epsfysize=25 cm \leavevmode \epsfbox{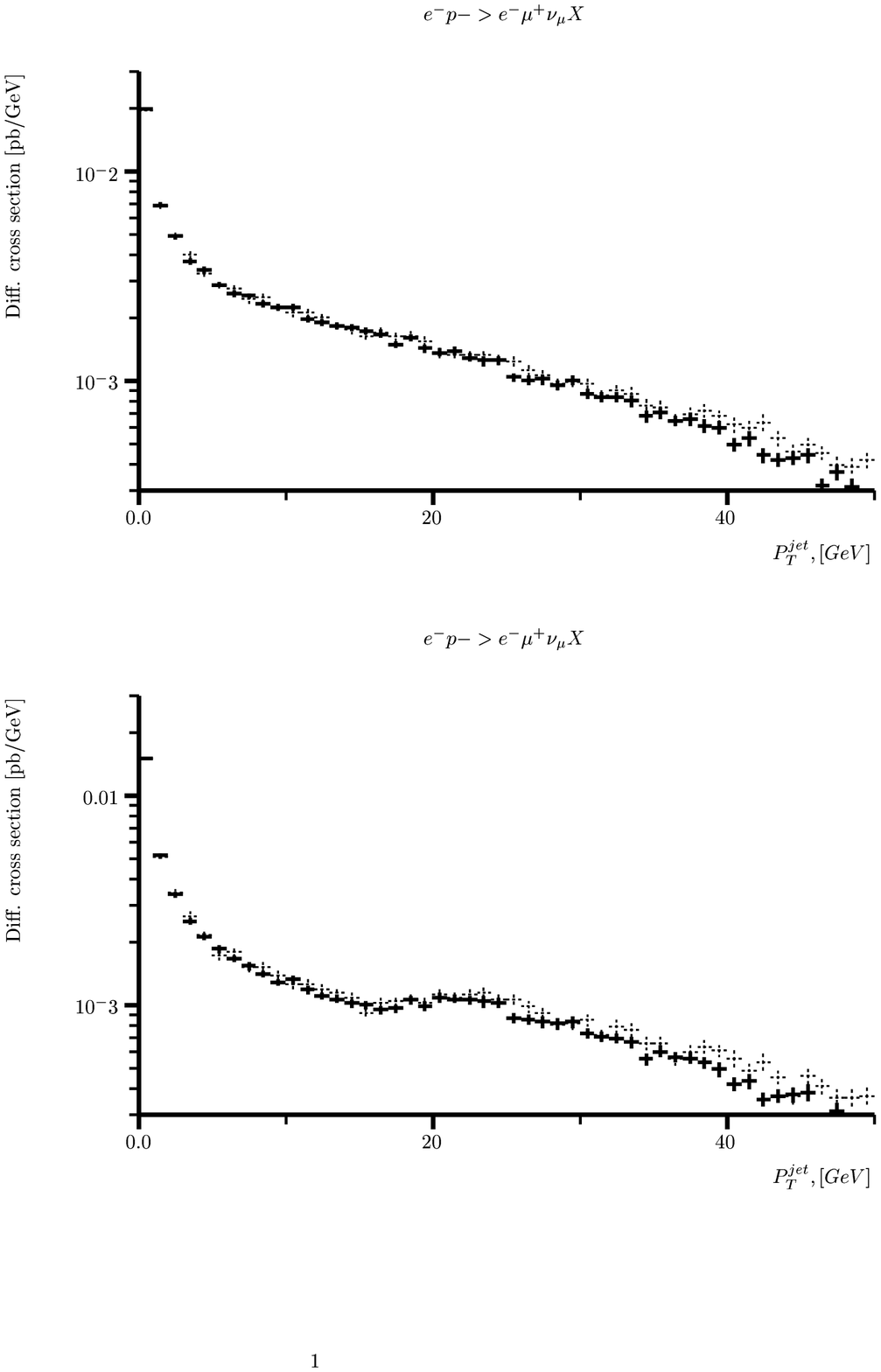}}
\end{picture}
\caption{Distribution of jet transverse momentum in the reaction
$e^- p \rightarrow e^- \mu^+ \nu_{\mu} X$. 
Upper plot: no kinematical cuts, solid lines - standard case, dash 
lines - anomalous three vector boson couplings case, $\lambda=1$, $k=1$ 
Lower plot: the same distributions after kinematical cuts
$E_{\mu} \ge$ 10 GeV, missing $p_T \ge$ 20 GeV.}
\end{figure}

\newpage

\begin{figure}
\begin{picture}(15,15)
\put(-6,-3){\epsfxsize=19cm
         \epsfysize=25 cm \leavevmode \epsfbox{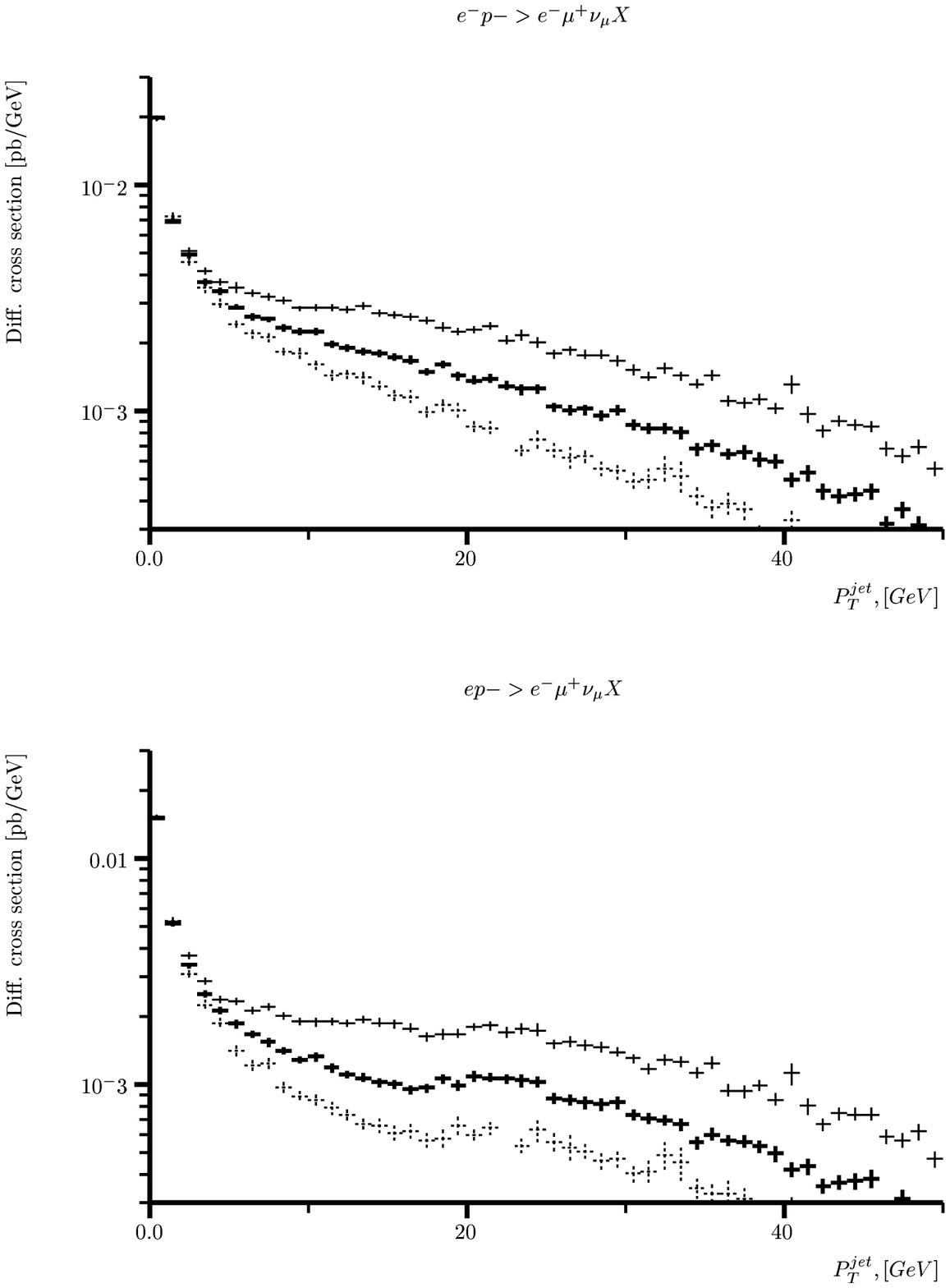}}
\end{picture}
\caption{Distribution of jet transverse momentum in the reaction
$e^- p \rightarrow e^- \mu^+ \nu_{\mu} X$.
Upper plot: no kinematical cuts, solid lines - standard case, dash
lines - anomalous three vector boson couplings case, $\lambda=0$, $k=0$,
thin solid lines - $\lambda=0$, $k=2$.
Lower plot: the same distributions after kinematical cuts
$E_{\mu} \ge$ 10 GeV, missing $p_T \ge$ 20 GeV.}
\end{figure}


\newpage

\begin{figure}
\begin{picture}(15,15)
\put(-6,-3){\epsfxsize=19cm
         \epsfysize=25 cm \leavevmode \epsfbox{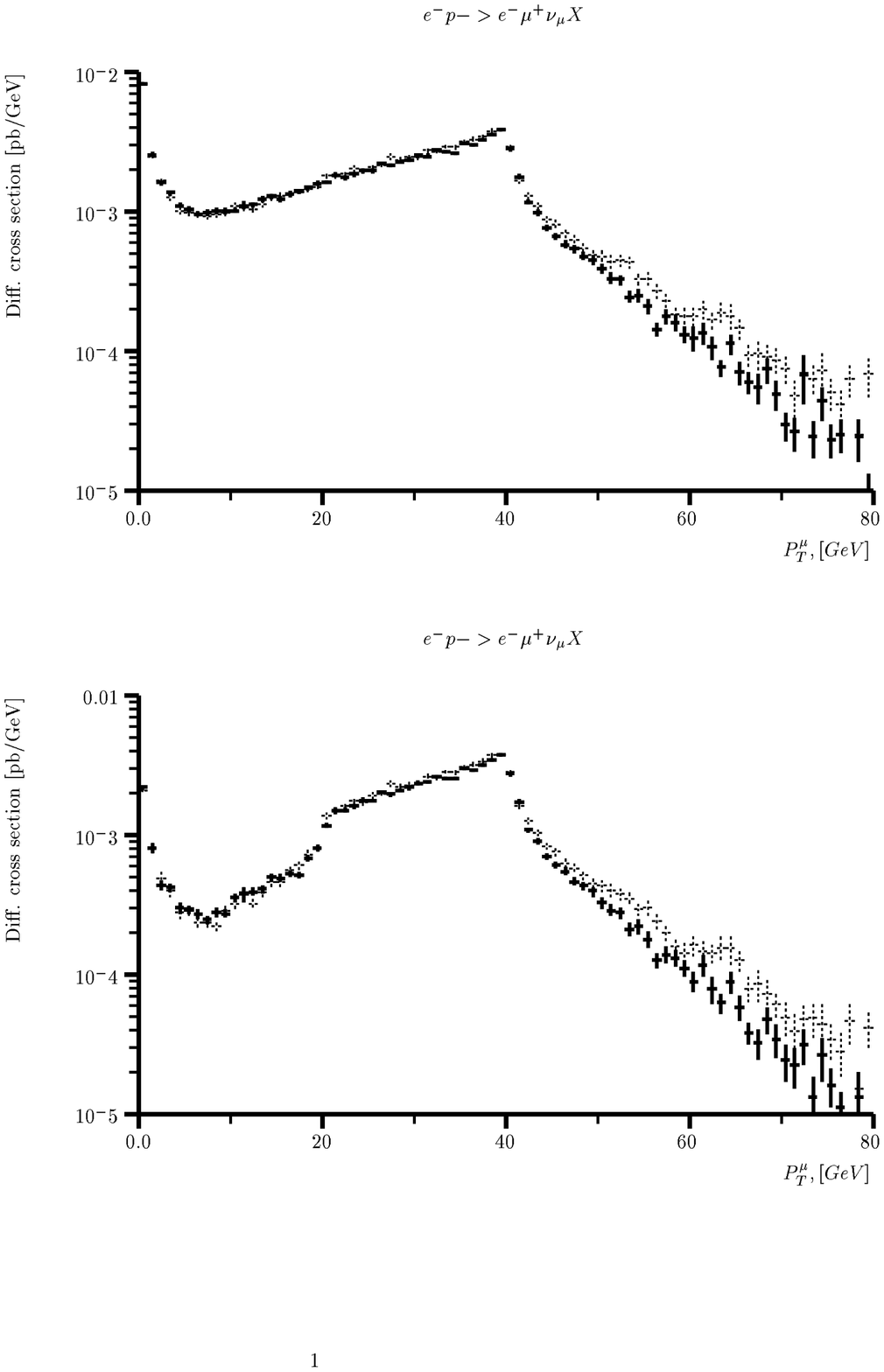}}
\end{picture}
\caption{Distribution of muon transverse momentum in the reaction
$e^- p \rightarrow e^- \mu^+ \nu_{\mu} X$.
Upper plot: no kinematical cuts, solid lines - standard case, dash
lines - anomalous three vector boson couplings case, $\lambda=1$, $k=1$
Lower plot: the same distributions after kinematical cuts
$E_{\mu} \ge$ 10 GeV, missing $p_T \ge$ 20 GeV.}
\end{figure}

\newpage

\begin{figure}
\begin{picture}(15,15)
\put(-6,-3){\epsfxsize=19cm
         \epsfysize=25 cm \leavevmode \epsfbox{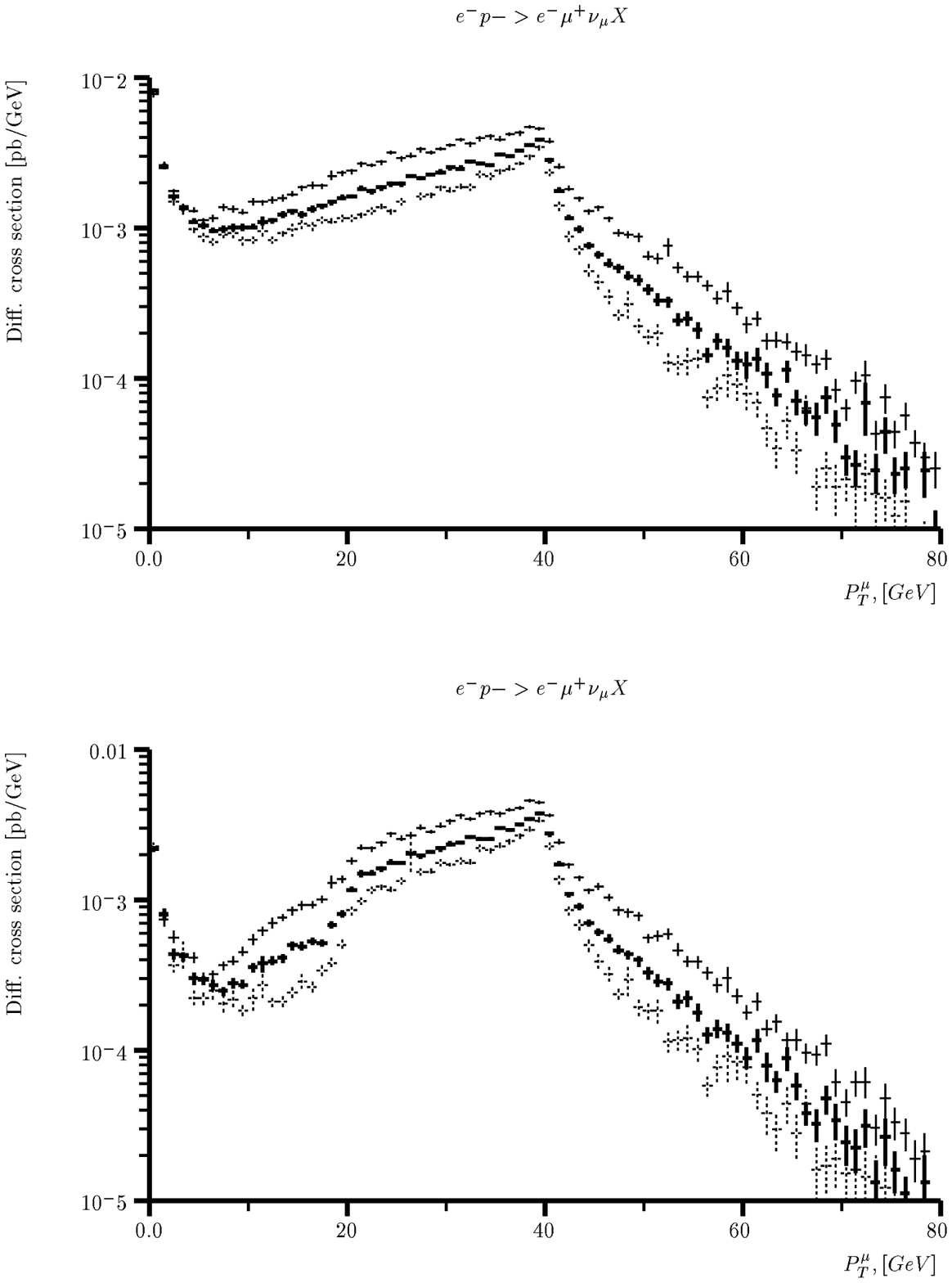}}
\end{picture}
\caption{Distribution of muon transverse momentum in the reaction
$e^- p \rightarrow e^- \mu^+ \nu_{\mu} X$.
Upper plot: no kinematical cuts, solid lines - standard case, dash
lines - anomalous three vector boson couplings case, $\lambda=0$, $k=0$,
thin solid lines - $\lambda=0$, $k=2$.
Lower plot: the same distributions after kinematical cuts
$E_{\mu} \ge$ 10 GeV, missing $p_T \ge$ 20 GeV.}
\end{figure}

\newpage

\begin{figure}
\begin{picture}(15,15)
\put(-6,-3){\epsfxsize=19cm
         \epsfysize=25 cm \leavevmode \epsfbox{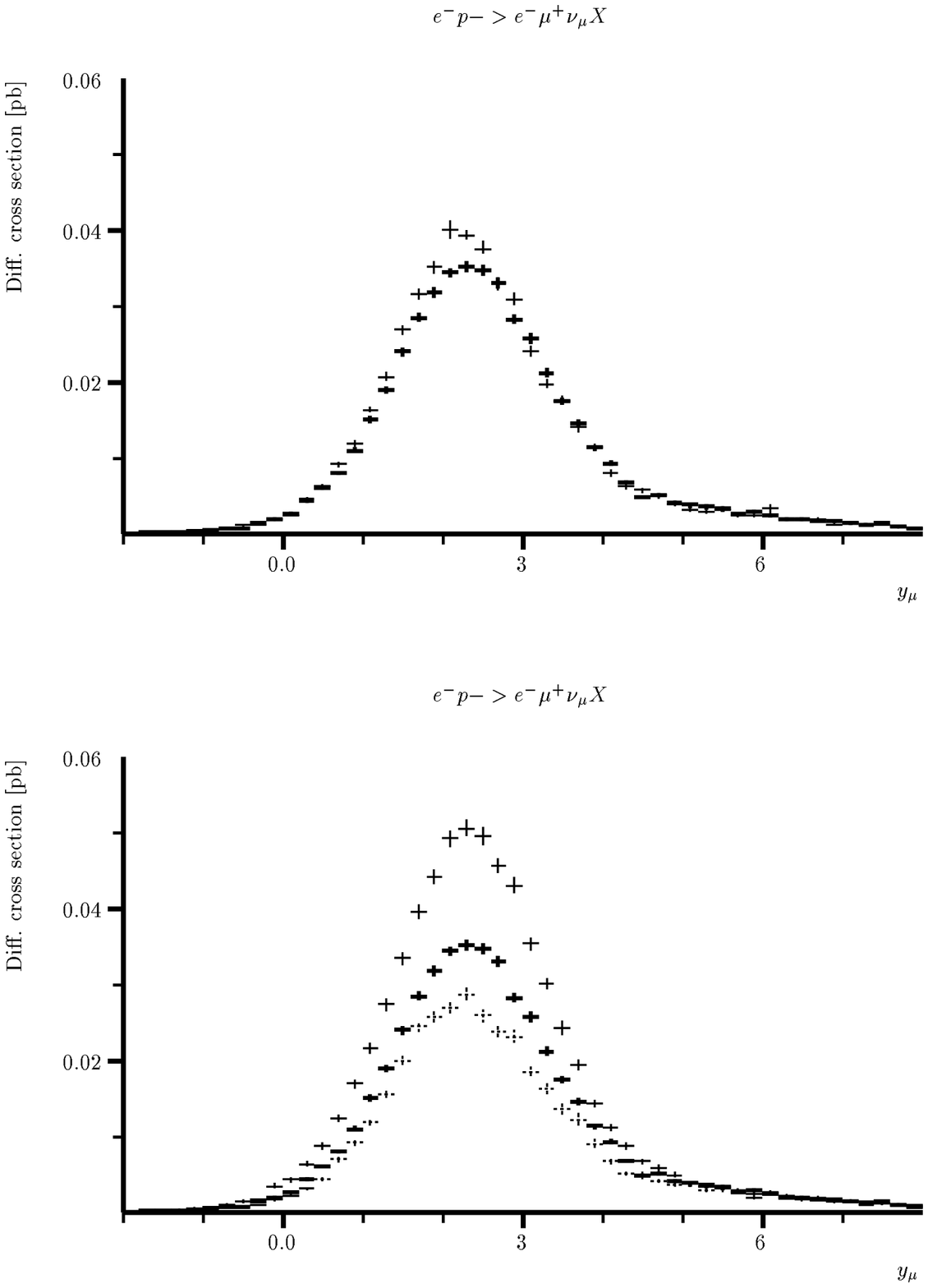}}
\end{picture}
\caption{Distribution of muon rapidity in the reaction
$e^- p \rightarrow e^- \mu^+ \nu_{\mu} X$.
Upper plot: no kinematical cuts, solid lines - standard case, dash
lines - anomalous three vector boson couplings case, $\lambda=1$, $k=1$.
Lower plot: no kinematical cuts, solid lines - standard case, dash
lines - anomalous three vector boson couplings case, $\lambda=0$, $k=0$,
thin solid lines - $\lambda=0$, $k=2$.}
\end{figure}

\end{document}